\documentclass[english]{revtex4}
\usepackage[T1]{fontenc}
\usepackage[latin1]{inputenc}

\makeatletter

\usepackage{babel}
\makeatother
\begin{document}

\title{Thermodynamical Analogy Between BTZ Black Holes and Effective String
Theory}

\author{Eduard Alexis Larrañaga Rubio}

\email{eduardalexis@gmail.com}

\affiliation{National University of Colombia}

\affiliation{National Astronomical Observatory (OAN)}

\begin{abstract}
In this paper we study the first law of thermodynamics for the (2+1)
dimensional BTZ rotating black hole considering a pair of thermodinamical
systems constructed with the two horizons of this solution. We show
that these two systems are similar to the right and left movers of
string theory and that the temperature associated with the black hole
is the harmonic mean of the temperatures associated with these two
systems.
\end{abstract}
\maketitle

\section{Introduction}

Bekenstein and Hawking showed that black holes have non-zero entropy
and that they emit a thermal radiation that is proportional to its
surface gravity at the horizon. These two quantities are related with
the mass through the identity

\begin{equation}
dM=TdS,\end{equation}

that is called \emph{the first law of black hole thermodynamics}\cite{hawking,bekenstein}.
But when the black hole has other properties as angular momentum $\mathbf{J}$
and electric charge $Q$, the first law is generalized to 

\begin{equation}
dM=TdS+\Omega dJ+\Phi dQ,\end{equation}

where $\Omega=\frac{\partial M}{\partial J}$ is the angular velocity
and $\Phi=\frac{\partial M}{\partial Q}$ is the electric potential.
When the black hole has two horizons, it is known that it is possible
to associate a first law with each of them. The outer horizon is related
with the Hawking radiation while the inner horizon is related with
the absortion proccess.\\

In this paper we will apply the method described by Wu \cite{wu2}
to describe the thermodynamics of the rorating BTZ black hole in (2+1)
dimensions and try to relate it with the effective string theory and
D-brane description of black holes..

In order to obtain this, we will define two thermodynamical systems
as the sum and the difference of the two horizons associated with
the rotating BTZ black hole. These systems resemble the R and L moving
modes of string theory. Then, we will describe the thermodynamics
of each of these systems to show how the Hawking temperature $T_{H}$
associated with the BTZ black hole can be interpreted as the harmonic
mean of thetemperature of the R and L parts, i.e.

\begin{equation}
\frac{2}{T_{H}}=\frac{1}{T_{R}}+\frac{1}{T_{L}}.\end{equation}

\section{The BTZ Rotating Black Hole}

The rotating BTZ black hole \cite{martinez} is a solution of $\left(2+1\right)$
dimensional gravity with a negative cosmological constant $\Lambda=-\frac{1}{l^{2}}$.
Its line element can be written as

\begin{equation}
ds^{2}=-\Delta dt^{2}+\frac{dr^{2}}{\Delta}+r^{2}d\varphi^{2},\end{equation}

where the lapse function is

\begin{equation}
\Delta=-M+\frac{r^{2}}{l^{2}}+\frac{J^{2}}{4r^{2}}.\end{equation}

This solution has two horizons given by the condition $\Delta=0$,

\begin{equation}
r_{\pm}^{2}=\frac{Ml^{2}}{2}\left[1\pm\sqrt{1-\frac{J^{2}}{M^{2}l^{2}}}\right].\label{eq:radii}\end{equation}
The mass of the black hole can be written in terms of this horizons
as

\begin{equation}
M=\frac{r_{+}^{2}+r_{-}^{2}}{l^{2}},\label{eq:mass1}\end{equation}

while the angular momentum $J$ is given by

\begin{equation}
J=\frac{2r_{+}r_{-}}{l}.\end{equation}

The Bekenstein-Hawking entropy associated with the black hole is twice
the perimeter of the outer horizon,

\begin{equation}
S=4\pi r_{+},\end{equation}
and therefore, the mass can be written as

\begin{equation}
M=\frac{S^{2}}{16\pi^{2}l^{2}}+\frac{4\pi^{2}J^{2}}{S^{2}}.\label{eq:massformula}\end{equation}

This expression can be re-written as \cite{wang}

\begin{equation}
M=\frac{1}{2}TS+\Omega J=\kappa\mathcal{P}+\Omega J,\label{eq:generalIntFirstLaw}\end{equation}

where $\mathcal{P}=\frac{P}{2\pi}$ is the {}``reduced'' perimeter
and $\kappa$ is the surface gravity. The Hawking-Bekenstein entropy
can be written as

\begin{equation}
S=4\pi\mathcal{P},\end{equation}
so the mass (\ref{eq:massformula}) is given by

\begin{equation}
M=\frac{\mathcal{P}^{2}}{l^{2}}+\frac{J^{2}}{4\mathcal{P}^{2}}.\end{equation}

Finally, the differential form of the first law for this black hole
takes the form \cite{Akbar}

\begin{equation}
dM=2\kappa d\mathcal{P}+\Omega dJ.\label{eq:generalDiffFirstLaw}\end{equation}

As is well known, we can associate a thermodynamics to the outer horizon
when treated as a single thermodynamical system. With the four laws
associated with this horizon one can describe the Hawking radiation
process. On the other hand, it is also possible to consider the inner
horizon as an independient thermodynamical system and associate it
a set of four laws that are related with the Hawking absorption process
\cite{Wu}. Therefore, the integral and differential mass formulae
can be written for the two horizons,

\begin{eqnarray}
M & = & \frac{\mathcal{P_{\pm}}^{2}}{l^{2}}+\frac{J^{2}}{4\mathcal{P_{\pm}}^{2}}\\
dM & = & 2\kappa_{\pm}d\mathcal{P}_{\pm}+\Omega_{\pm}dJ.\end{eqnarray}
From these relations, is easy to see that the surface gravity and
angular velocity at the two horizons are

\begin{eqnarray}
\kappa_{\pm} & = & \frac{1}{2}\left.\frac{\partial M}{\partial\mathcal{P}_{\pm}}\right|_{J}=\frac{\mathcal{P_{\pm}}}{l^{2}}-\frac{J^{2}}{4\mathcal{P_{\pm}}^{3}}=\frac{r_{\pm}}{l^{2}}-\frac{J^{2}}{4r_{\pm}^{3}}\\
\Omega_{\pm} & = & \left.\frac{\partial M}{\partial J}\right|_{\mathcal{P}_{\pm}}=\frac{J}{2\mathcal{P_{\pm}}^{2}}=\frac{J}{2r_{\pm}^{2}},\end{eqnarray}

while the entropy and temperature associated with each horizon are

\begin{eqnarray}
S_{\pm} & = & 4\pi\mathcal{P}_{\pm}\\
T_{\pm} & = & \frac{\kappa_{\pm}}{2\pi}.\end{eqnarray}

Using the inner and outer horizons we will define two independient
thermodynamical systems. Following Wu \cite{wu2}, the R-system will
have a reduced perimeter corrspondient to the sum of the inner and
outer perimeters while L-system corresponds to the difference of these
perimeters,

\begin{eqnarray}
\mathcal{P}_{R} & = & \mathcal{P}_{+}+\mathcal{P}_{-}\\
\mathcal{P}_{L} & = & \mathcal{P}_{+}-\mathcal{P}_{-}.\end{eqnarray}

It is important to note that each of these systems carry two hairs
$\left(M,J\right)$, but we will show that they represent two black
holes with an asymmetry in the angular momentum. However, to begin,
we will obtain the thermodynamical relations for these systems and
after that we will relate them with the thermodynamics of the BTZ
black hole and its Hawking temperature.

\section{R-System Thermodynamics}

First, we will focus in the R-system. Using the expression for the
inner and outer radii (\ref{eq:radii}) we can write

\begin{equation}
\mathcal{P}_{R}^{2}=\mathcal{P}_{+}^{2}+\mathcal{P}_{-}^{2}+Jl=Ml^{2}+Jl.\end{equation}
Therefore, the mass formula (\ref{eq:mass1}) can be written as

\begin{equation}
M=\frac{\mathcal{P}_{R}^{2}}{l^{2}}-\frac{J}{l}.\end{equation}

The surface gravity and the angular velocity associated with this
system are given by

\begin{eqnarray}
\kappa_{R} & = & \frac{1}{2}\left.\frac{\partial M}{\partial\mathcal{P}_{R}}\right|_{J}=\frac{\mathcal{P}_{R}}{l^{2}}\\
\Omega_{R} & = & \left.\frac{\partial M}{\partial J}\right|_{\mathcal{P}_{R}}=-\frac{1}{l},\end{eqnarray}

and the correspondient entropy and tempertature are

\begin{eqnarray}
S_{R} & = & 4\pi\mathcal{P}_{R}=4\pi\left(\mathcal{P}_{+}+\mathcal{P}_{-}\right)=S_{+}+S_{-}\label{eq:entropyR}\\
S_{R} & = & 4\pi\sqrt{Ml^{2}+Jl}\end{eqnarray}

\begin{eqnarray}
T_{R} & = & \frac{\kappa_{R}}{2\pi}=\frac{\mathcal{P}_{R}}{2\pi l^{2}}=\frac{\sqrt{Ml^{2}+Jl}}{2\pi l^{2}}.\end{eqnarray}

Then, the integral and differential mass formulae for the R-system
are

\begin{eqnarray}
M & = & \kappa_{R}\mathcal{P}_{R}+\Omega_{R}J\\
dM & = & \frac{2\mathcal{P}_{R}}{l^{2}}d\mathcal{P}_{R}-\frac{1}{l}dJ=2\kappa_{R}d\mathcal{P}_{R}+\Omega_{R}dJ,\end{eqnarray}
that corresponds to what is expected from equations (\ref{eq:generalIntFirstLaw})
and (\ref{eq:generalDiffFirstLaw}).

\section{L-System Thermodynamics}

Now, we will turn our attention to the L-system. Using again the expression
(\ref{eq:radii}) we have

\begin{equation}
\mathcal{P}_{L}^{2}=\mathcal{P}_{+}^{2}+\mathcal{P}_{-}^{2}-Jl=Ml^{2}-Jl.\end{equation}
Therefore, the mass formula (\ref{eq:mass1}) can be written now as

\begin{equation}
M=\frac{\mathcal{P}_{L}^{2}}{l^{2}}+\frac{J}{l}.\end{equation}

The surface gravity and the angular velocity associated with the L-system
are given by

\begin{eqnarray}
\kappa_{L} & = & \frac{1}{2}\left.\frac{\partial M}{\partial\mathcal{P}_{L}}\right|_{J}=\frac{\mathcal{P}_{L}}{l^{2}}\\
\Omega_{L} & = & \left.\frac{\partial M}{\partial J}\right|_{\mathcal{P}_{L}}=\frac{1}{l},\end{eqnarray}

and the entropy and tempertature are now

\begin{eqnarray}
S_{L} & = & 4\pi\mathcal{P}_{L}=4\pi\left(\mathcal{P}_{+}-\mathcal{P}_{-}\right)=S_{+}-S_{-}\label{eq:entropyL}\\
S_{L} & = & 4\pi\sqrt{Ml^{2}-Jl}\end{eqnarray}

\begin{eqnarray}
T_{L} & = & \frac{\kappa_{L}}{2\pi}=\frac{\mathcal{P}_{L}}{2\pi l^{2}}=\frac{\sqrt{Ml^{2}-Jl}}{2\pi l^{2}}\label{eq:Tl}\end{eqnarray}

The integral and differential mass formulae for this system are

\begin{eqnarray}
M & = & \kappa_{L}\mathcal{P}_{L}+\Omega_{L}J\\
dM & = & \frac{2\mathcal{P}_{L}}{l^{2}}d\mathcal{P}_{L}-\frac{1}{l}dJ=2\kappa_{L}d\mathcal{P}_{L}+\Omega_{L}dJ,\end{eqnarray}
that corresponds again to equations (\ref{eq:generalIntFirstLaw})
and (\ref{eq:generalDiffFirstLaw}). It is important to note here
that equation \ref{eq:Tl} lets identify the extremal black hole case
(under the condition $Ml=J$), that leave us only with one system
( i.e. the R-system).

\section{Relationship between the R,L-systems and the BTZ thermodynamics}

The thermodynamic laws of the R, L- systems are related with the BTZ
black hole thermodynamics. To see this, note that the surface gravities
of the R and L systems can be written in terms of the surface gravity
of the two horizons as 

\begin{equation}
\frac{1}{\kappa_{R,L}}=\frac{l^{2}}{\mathcal{P}_{R,L}}=\frac{1}{\kappa_{+}}\pm\frac{1}{\kappa_{-}},\end{equation}
that corresponds exactly with the relation found by Wu\cite{wu2}
for Ker-Newman black hole, and that is in direct correspondence to
string theory and D-brane physics. Since temperature is proportional
to surface gravity, we have a similar expression,

\begin{equation}
\frac{1}{T_{R,L}}=\frac{2\pi l^{2}}{\mathcal{P}_{R,L}}=\frac{1}{T_{+}}\pm\frac{1}{T_{-}}.\label{eq:temperatures}\end{equation}

Note that from this relation is immediate to obtain the expression
for the Hawking temperature associated with the BTZ black hole, that
corresponds to the temperature of the outer horizon,

\begin{equation}
T_{H}=T_{+}=\frac{\kappa_{+}}{2\pi}.\end{equation}

Using equation (\ref{eq:temperatures}), the Hawking temperature can
be written as the harmonic mean of the R and L temperatures,

\begin{equation}
\frac{2}{T_{H}}=\frac{1}{T_{R}}+\frac{1}{T_{L}}.\end{equation}

It is also important to note that there is an asymmetry between the
angular momenta of the R and L systems as can be seen from the angular
velocities,

\begin{equation}
\Omega_{R}=-\frac{1}{l}=-\Omega_{L}.\end{equation}

Finally, note that in the extremal case, given by the condition $Ml=J$,
the R and L systems are indistinguishable. This means that they are
in thermal equilibrium with a temperature

\begin{equation}
T_{E}=\frac{\sqrt{2Jl}}{2\pi l^{2}}=\frac{r_{+}^{2}}{\pi l^{2}}.\end{equation}

\section{conclusion}

In this paper we have shown that the thermodynamics of the (2+1) dimensional
rotating BTZ black hole can be constructed from two independient thermodynamical
systems constructed from the two horizons and that resemble the right
and left modes of string theory. If one assume that the effective
strings have the same mass and angular momentum that the rotating
BTZ black hole, there is a correspondence between the R and L modes
thermodynamics and the thermodynamics of the horizons. 

For example, note that the entropy of the R and L systemes are obtained
as the sum and difference of the entropies of the two horizons (equations
\ref{eq:entropyR}-\ref{eq:entropyL}), as in effective string theory.
Moreover, we have show that the Hawking temperature associated with
the black hole is obtained as the harmonic mean of the temperatures
associated with the R and L systems, just as in the case of stringy
thermodynamics. Thus, the picture of black holes provided by effective
strings and D-Branes seem to hold in (2+1) gravity, even for entropy
calculation. \\

All these facts suggest that, as proposed by Wu\cite{wu2}, there
is a deep connection between string theory and D-branes with black
holes physics, and this relation can lead to the understanding of
the origin of entropy, not only in general relativity but also in
other dimensions.

\end{document}